# Définition d'une structure adaptative de réseau local sans fil à consommation optimisée

Sabri CHEBIRA, Gilles Mercier et Jackson Francomme

*LIIA-120-122, rue Paul Armangot 94400 Vitry Sur Seine-
Université de Créteil (France)*

**ms.chebira@univ-paris12.fr**
**mercier@univ-paris12.fr**
**francomme@univ-paris12.fr**

**Résumé:** La forte croissance des réseaux sans fil bas débit (LR-WPAN), nous amène à considérer les problèmes d'autonomie, donc de durée de vie des nœuds d'un réseau, sachant que le changement des alimentations est souvent difficile à effectuer, d'une part. La mobilité inhérente à ce type de matériel est un élément essentiel, il en résultera des contraintes de routage, donc un problème complexe à résoudre, d'autre part. Cet article propose des axes de travail pour évaluer les performances d'un tel réseau en terme de consommation énergétique et en performance de mobilité. Les objectifs étant contradictoires, il faudra trouver nécessairement un compromis. Par ailleurs, si nous souhaitons garantir un délai maximal de remise de message, possibilité que nous offre la norme 802.15.4, un autre compromis s'imposera entre une structure strictement figée, et une structure totalement mobile. Nous présentons donc, une quantification du coût énergétique en fonction d'un débit souhaité, par rapport à la durée de sommeil des nœuds du réseau. Puis nous ouvrons des pistes de réflexion pour trouver le meilleur compromis : consommation/mobilité/délais garantis, en suggérant une structure de réseau adaptative partant d'un concept de MANET.

**Mots clés:** Autonomie, Consommation, Délais garantis, Mobilité, Réseau Local sans fil (LR-WLAN), Routage, MANET, IEEE 802.15.4.

## 1 Introduction

Il semble très vraisemblable, que les réseaux sans fil bas débit LR-WPAN (Low Rate Wireless Personal Area Network) prendront de plus en plus de place dans notre vie quotidienne et dans une gamme d'applications très variées, incluant la domotique, l'industrie des procédés, etc.. Pour ces types d'applications, un nombre fini de nœuds mobiles et autonomes est distribué sur une surface et communiquent sans fils. Comparé au WLAN (Wireless Local Area Network) qui sont conçus pour fournir un débit plus important et une faible latence pour des environnements de transfert de fichiers et pour le multimédia, le débit requis par les applications sous un réseau LR-WPAN reste de l'ordre de quelques dizaines de kbits/sec. En conséquence, la latence requise est de l'ordre des 100ms ou plus (Gallaway & al, 2002). Le paramètre le plus important dans ces applications est la consommation en énergie électrique qui doit être la plus bas possible, puisqu' en pratique, il est difficile de changer les batteries ou de les recharger dans de nombreux cas.

Un nouveau standard à été développé avec ces objectifs par le groupe de travail IEEE 802.15.4 (Specifications 802.15.4-2003 IEEE) (Gutierrez & al, 2001). Le but de cette technologie est de fournir un standard pour les couches basses du modèle OSI (physique et liaison de données) offrant un bas débit et une consommation d'énergie faible et modulable.

Cet article présente une analyse des facteurs d'influence sur la consommation énergétique du réseau, donc dans l'autonomie des noeuds qui le constituent. Cette analyse portera essentiellement sur les aspects de débit, de garantie des délais de transmission de message, de mobilité et d'évolutivité du réseau. Sachant que ces critères de performances ont des effets contradictoires sur les performances énergétiques, nous chercherons à élaborer des compromis pour un choix optimal dans une application donnée.

Au paragraphe 2, nous présentons les principales caractéristiques de la couche MAC du IEEE 802.15.4.

Le paragraphe 3 expose une analyse des facteurs d'influence sur les performances du IEEE 802.15.4 (débit/ consommation de ce standard)

Au paragraphe 4 nous présentons un exemple de réseau adaptatif et analysons les caractéristiques d'adaptation de ce concept au standard 802.15.4

## 2. Norme IEEE 802.15.4

La norme IEEE 802.15.4 définit deux modes d'accès au média. Le premier est le mode sans balises, que l'on ne considérera pas dans cet article parce qu'il est rarement utilisé, et qu'il ne peut répondre aux objectifs de garantie de délais d'acheminement des messages. Le second est le mode balisé. C'est sur ce mode que nous développons notre analyse.

Dans le mode balisé, la norme IEEE 802.15.4 utilise une structure de super trame. La super trame commence par une balise transmise périodiquement, par un nœud spécifique du réseau appelé coordinateur, à intervalles de temps réguliers de l'ordre de 15ms à 245s. Entre deux balises se définit la super trame, dans laquelle cohabitent deux périodes :

- une période **active** où les nœuds peuvent communiquer.

- une période **non active** dans la quelle les nœuds se mettent en mode veille.

Un paramètre permet de fixer la durée des intervalles des balises. Ce paramètre est désigné par le terme (*macBeaconOrder : BO*) ($BO=2^{BO} \times aBaseSuperFrameDuration$). Un autre paramètre (*macSuperFrameOrder : SO*) définit la longueur de la portion active de la super trame. La partie active de la super trame est divisée en 16 slots de temps égaux et consiste en trois parties : la balise, la période d'accès par contention (CAP : Contention Access Period) et la période de libre collision (CFP : Contention Free Period). Dans le cas où la super trame inclue un segment CFP, le coordinateur garantit l'accès au médium à certains nœuds. Pour cela, le segment CFP est divisé en plusieurs parties appelées GTS (Guaranteed time slots). Chaque GTS consiste en de multiples entiers de slots CFP et il peut y avoir jusqu'à 7 GTS dans une CFP. Nous présentons maintenant quelques particularités de la couche MAC de la norme IEEE 802.15.4. (Specifications 802.15.4-2003 IEEE)

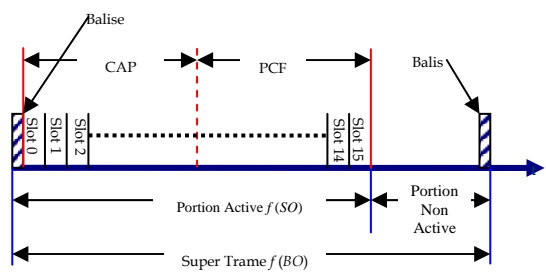

**Figure 1.** *Structure d'une super trame 802.15.4*

### 2.1 Transmission des données :

Avant toute transmission de données, les nœuds ont besoin de connaître la structure de la super trame. Pour cela ils doivent attendre la transmission da la balise. Si un nœud a réservé un créneau GTS, il envoie les données durant le segment CFP, sinon il les envoie pendant le segment CAP.

Dans le PAN (Personal Area Network), le coordinateur du PAN est responsable de la génération des balises super trames et de toutes les information la composant (*SO, BO,* les GTSs, …)

En ce qui concerne les communications du coordinateur en destination des nœuds La requête de transaction doit être initialisée par le nœud lui-même afin de permettre à des dispositifs d'être en mode d'économie d'énergie selon leur propre volonté.

Le nœud envoie une commande de requête de données au coordinateur pendant le segment CAP. Si son adresse est dans la liste d'attente de la balise, le coordinateur envoie une trame d'acquittement avec un drapeau indiquant que les données seront envoyées prochainement. Il envoie ensuite la trame. Quand les données sont reçues, un acquittement est envoyé au coordinateur du PAN.

### 2.2 Rapport cyclique

Le coordinateur annonce la structure de la super trame aux nœuds du PAN périodiquement à travers les trames balise. En changeant les portions actives et non actives via les paramètres SO et BO, le coordinateur doit opérer sous un faible rapport cyclique afin d'économiser l'énergie

Le rapport cyclique étant égal à $2^{SO}/2^{BO}$.

### 2.3 La période d'accès par contention (CAP)

Le segment CAP est situé entre la balise de la super trame et le segment CFP. Toutes les trames dans le segment CAP utilisent la technique d'accès au média de type Slotted CSMA-CA comme décrit ci-dessous :

Trois variables sont mises à jour dans chaque nœud pour l'accès au canal : NB, CW et BE. NB est la période du BackOff dans le CSMA-CA attendant la fin de la transmission en cours, et reste à zéro pour chaque nouvelle transmission. CW est la longueur de la fenêtre de contention, qui est à 2 lors d'une nouvelle transmission de données ou bien quand le canal est occupé. BE est l'exposant BackOff, qui est relié à la période du BackOff que le nœud doit attendre avant la prochaine tentative d'accès au canal (écoute du canal).

Quand le nœud a besoin de transmettre des paquets pendant le segment CAP, il initialise sa réception et attend un nombre aléatoire de périodes de BackOff complètes (jusqu'à $2^{BE}-1$ périodes) et détermine si le canal est libre ou pas. La couche MAC assure qu'après le BackOff aléatoire, les opérations CSMA-CA restantes seront prises en compte et la transaction entière sera complétée avant la fin du segment CAP. Si le canal est occupé, on incrémente NB et BE de un, et on réinitialise CW à 2. Si NB est plus petit ou égale qu'un seuil désigné par le terme macCSMABackOffs, on attend pendant une autre période aléatoire, sinon la tentative de transmission se termine par un échec.

Cette technique d'accès au medium conduit à une augmentation de la consommation d'énergie dans le cas où on a de longues périodes de BackOff. Ce problème apparaît lors de périodes à grand trafic pour éviter les collisions. Cependant, la norme IEEE 802.15.4 supporte le mode d'extension de vie de batterie (BLE : Battery Life Extention), dans lequel les composantes BackOff sont limitées entre 0 et 2. Ce dispositif réduit les périodes d'écoute du canal libre pour les applications à faible trafic. Un nœud du réseau doit mettre sa radio en veille pour conserver l'énergie immédiatement après la réception de la trame d'acquittement s'il n'a pas d'autres données à transmettre ou à recevoir

### 2.4 La période d'accès garantis CFP

Le standard IEEE 802.15.4 permet l'utilisation optionnelle de segments CFP pour les nœuds demandant une bande passante propre pour avoir une faible latence. Le segment CFP se positionne entre la fin du segment CAP et le début de la prochaine balise. Toutes les transactions avec contention doivent se terminer avant le début du segment CFP.

Un nœud nécessitant une bande passante propre ou bien une faible latence doit se faire assigner par le coordinateur un GTS dans la CFP. Quand le nœud veut transmettre une trame à travers un créneau GTS. Il doit en premier lieu consulter la liste dans la trame balise pour connaître la position de son créneau GTS dans la super trame. Si un créneau GTS valide lui est fourni, le nœud se tient prêt à le localiser et transmet ses données durant cette fenêtre temporelle GTS.

### 2.5 La synchronisation

Le coordinateur du PAN, transmet périodiquement des trames « balise » pour annoncer la structure de la super trame dans le PAN. Les nœuds du PAN ont besoin de se synchroniser avec le coordinateur en recevant et décodant la trame balise avant le début des transmissions. Il existe deux méthodes de synchronisation : **avec suivi** et **sans suivi**.

Dans le cas d'une synchronisation avec suivi, le nœud reçoit en premier la balise, reconnaît la structure de la super trame, reconnaît le moment ou il doit activer une réception pour la prochaine balise et faire un suivi. Pour transmettre une trame, le nœud doit activer la réception un peu avant l'arrivée de la balise.

Dans le cas d'une synchronisation sans suivi, Si le nœud à besoin de transmettre des données, il essaie d'acquérir la balise sans qu'il sache à quel moment elle sera envoyée. Ainsi le nœud est obligé d'activer la réception jusqu'à ce qu'il reçoive une balise de la part de son coordinateur.

## 3. Analyse des facteurs d'influences sur les performances du réseau

Dans cette partie, nous exposerons les différents facteurs qui influent sur les performances du réseau.

Ses performances sont exprimées en terme de bande passante, de latence et de consommation et dépendent des besoins de l'application. La norme IEEE 802.15.4 met à disposition un certain nombre de fonctionnalités paramétrables, que nous nous proposons maintenant d'analyser:

### 3.1 Le CSMA-CA dans la CAP

Dans la norme IEEE 802.15.4, la technique d'accès au média (CSMA-CA) emploie de courtes périodes de BackOff afin de réduire la consommation d'énergie nécessaire à l'écoute du média libre. Ceci n'est vraiment réaliste que pour des applications à temps non critique ou lentes.

Les délimiteurs de la super trame qui sont les balises, ne permettent pas à un nœud d'émettre si le temps nécessaire pour la transmission est plus grand que la durée du segment CAP. Ceci se traduit en une moindre efficacité de l'utilisation du média. Donc il faut paramétrer les balises pour avoir un segment CAP dont la durée peut contenir ce message.

### 3.2 Le rapport cyclique

Dans la norme IEEE 802.15.4 c'est le coordinateur qui décide du rapport cyclique et des intervalles de balises. Ceci est fait en fixant les paramètres BO et SO.

Un rapport cyclique bas conserve l'énergie en augmentant la période d'inactivité (nœuds en mode sommeil) mais cela réduit la bande passante et augmente la latence. Avec le même rapport cyclique on peut avoir différentes combinaisons de BO et SO ce qui veut dire qu'on aura différentes structure de la super trame au niveau des périodes active et inactive et au niveau de la longueur de la super trame. Ce qui revient à dire que lorsque le rapport cyclique diminue la consommation en énergie diminue (Linéairement). La réception des balises consomme la plupart de l'énergie tant que le coût en transmission est négligeable.

Le délai de délivrance des paquets augmente avec la diminution du rapport cyclique puisque l'on aura une grande période d'inactivité.

Avec le même rapport cyclique, le plus petit *SO* correspond à une période active plus petite, et puisque la durée de la balise est plus faible, cela conduit à une plus grande consommation : pour une période fixe, on aura plus de balises transmises et reçues. En contrepartie, on obtient une latence plus petite puisque si un paquet arrive dans la période inactive il aura une plus petite période d'attente.

### 3.3 Les créneaux GTS (Guaranteed Time Slots)

Les GTS sont assignés pendant le segment CFP aux applications demandant une bande passante dédiée ou une petite latence. Cependant, les créneaux de temps assignés à une application seront gaspillés s'il n y a pas de demande de transmission de données. De plus un nœud en mode GTS, doit décoder périodiquement les trames balise du coordinateur afin de localiser dans le temps le créneau qui lui est attribué et par la suite transmettre ses données. Ce qui implique que dans ce mode, le taux de délivrance soit de 100% et le noeud aura une faible latence par rapport à la charge du trafic dans le réseau. Cependant la consommation en énergie augmente

### 3.4 La synchronisation

Avant que le nœud ne transmette n'importe quelle trame de données, il a besoin de se synchroniser avec le coordinateur en recevant des trames balises. Avec le mode suivi de balises, le nœud active périodiquement sa radio afin de recevoir la trame balise, donc il consomme de l'énergie. Cependant, le nœud connaît le moment de l'émission de la prochaine trame balise, donc il se met en mode sommeil en attendant ce moment, afin de conserver de l'énergie.

A l'opposé, dans le mode sans suivi de balises, le nœud ne gaspille pas de l'énergie dans la réception périodique des trames balise. Cependant, tant qu'il ne sait pas à quel moment la trame balise est transmise, il doit mettre en marche sa radio dès qu'il aura besoin de communiquer avec le coordinateur. Ce qui peut coûter beaucoup en consommation d'énergie due à une longue période libre.

Il y a donc *un compromis à adopter entre les deux modes* : suivi et le non suivi des balises dépendant du rapport cyclique et du débit.

Pour une période fixe, Si le rapport cyclique est petit, le nombre de trames balise est réduit, donc dans le mode avec suivi de trame la réception des trames balise consomme moins d'énergie. En attendant les nœuds en mode non suivi de balise, (statistiquement) doivent attendre plus longtemps avant de recevoir une trame balise, donc ils doivent activer leurs récepteurs pendant une plus longue période, ce qui augmente la consommation en énergie. Avec un plus grand débit, la probabilité du besoin de transmettre des données dans l'intervalle balise augmente. En même temps il y reste un besoin de synchronisation pour les nœuds en mode non suivi de balise, et donc la consommation d'énergie augmente

*Analyse du coût énergétique des deux modes (avec et sans suivi)*

Nous étudions ici l'influence du choix du mode, avec et sans suivi, sur la consommation en énergie d'un nœud.

Soit $P_t$, $P_r$ et $P_i$ les puissances de consommation de la transmission radio, de la réception radio et de l'écoute libre. Soit $T_b$, $T_d$, $T_a$ et $T_i$ la durée en temps de la trame balise, trame de donnée, trame d'acquittement et le BackOff aléatoire respectivement. Soit $p$, la probabilité de la présence d'un paquet à transmettre dans l'intervalle balise BO. Supposons que le débit est r et K la taille du paquet à transmettre.

$$P = r \bullet \frac{BO}{K} \bullet 8$$

Pour le bas débit des WPAN, on suppose qu'il y a au plus une trame de donnée dans l'intervalle balise.

Avec le mode avec suivi de balise, le coût en énergie dans un intervalle balise de la super trame consiste en la réception d'une trame balise et s'il y a une transmission de trame, la consommation en énergie de la période de BackOff, la transmission de la trame de données et la réception de la trame d'acquittement. Ce qui donne :

$$E_{suivi} = P_r T_b + p(P_t T_d + P_r T_a + P_i T_i)$$

Sans suivi, le coût en énergie dans l'intervalle balise de la super trame est nul s'il n'y a pas de trame à transmettre. S'il y a des paquets à transmettre le nœud à besoin d'écouter le média pendant *BO/2* en moyenne dans le but de se synchroniser avec le coordinateur, ensuite il envoie le paquet dans le prochain intervalle balise. On a donc :

$$E_{Non-Suivi} = p\left(P_i + \frac{BO}{2} + P_t T_d + P_r T_a + P_i T_i\right)$$

## 4. Recherche d'une structure adaptative

La couche MAC de la norme 802.15.4 offre des services à la couche réseau. Afin de mieux comprendre les besoins et les contraintes propres à la couche réseau et de mieux comprendre le comportement de la couche réseau dans diverse topologie et d'applications, nous présentons un concept d'architecture de couche réseau pour les infrastructures mobiles ad hoc (MANET Mobile Ad hoc NETwork) basé grappe. Puis nous étudierons les……..

### 4.1 Caractéristiques des MANETs

Un réseau de type MANET (IETF, 2004) est constitué de plates-formes mobiles (i.e. un routeur avec plusieurs hôtes et éléments de communication sans fil). Actuellement appelés nSuds (nœuds) - ils sont libres de se déplacer arbitrairement. Ces nœuds peuvent être dans des avions, des bateaux, des camions, des voitures, voire même sur des personnes ou sur des éléments extrêmement petits, éventuellement avec plusieurs hôtes par routeur. Une structure MANET constitue un système autonome de nœuds mobiles. Ce système peut être isolé, mais il peut aussi avoir des passerelles ou des interfaces le reliant à un réseau fixe. Dans son fonctionnement futur, on le voie typiquement comme un réseau "final" (stub network) rattaché à un réseau d'interconnexion. Un réseau final gère le trafic créé ou à destination des nœuds internes, mais ne permet pas à un trafic extérieur de transiter par lui.

Les nœuds des MANET sont équipés d'émetteurs et de récepteurs sans fil utilisant des antennes qui peuvent être omnidirectionnelles (diffusion), fortement directionnelles (point à point), probablement orientables, ou une combinaison des deux. A un instant donné, en fonction de la position des nœuds, de la configuration de leur émetteur-récepteur, des niveaux de puissance de transmission et d'interférence entre les canaux, il y a une connectivité sans fil qui existe entre les nœuds, sous forme de graphe aléatoire multi sauts ou de réseau ad hoc. Cette topologie peut changer avec le temps en fonction du mouvement des nœuds ou de l'ajustement de leurs paramètres d'émission réception.

Les principales caractéristiques des MANETs sont les suivantes:

1- Topologies dynamiques : Les nœuds sont libres de se déplacer arbitrairement, ce qui fait que la topologie du réseau - typiquement multi sauts - peut changer aléatoirement et rapidement n'importe quand, et peut être constituée à la fois de liaisons unidirectionnelles et bidirectionnelles.

2- Liaisons à débits variables et à bande passante limitée : les liaisons sans fil auront toujours une capacité inférieure à leurs homologues câblés. En plus, le débit réel des communications sans fil - après avoir déduit les effets des accès multiples, du fading, du bruit, des interférences, etc. - est souvent inférieur aux taux de transfert maximum de la radio. Un des effets de ces débits de liaison relativement faibles est que la congestion sera généralement la norme plus que l'exception, i.e. La demande des applications distribuées approchera ou dépassera souvent la capacité du réseau. Comme le réseau mobile est souvent une simple extension d'un réseau fixe, les utilisateurs mobiles ad hoc demanderont les mêmes services. Cette demande ne cessera de croître avec l'augmentation des traitements multimédia et des applications basées sur les réseaux.

3- Utilisation limitée de l'énergie : Une partie des nœuds d'un MANET, voire l'ensemble des nœuds, peut reposer sur des bateries ou un autre moyen limité pour puiser leur énergie. Pour ces nœuds, le plus important est sans doute de mettre en place des critères d'optimisation pour la conservation de l'énergie.

Que peut on alors faire au niveau de la couche MAC afin d'optimiser le trafic sur le réseau ? Il est certain que la couche MAC ne peut pas à elle seule optimiser le trafic au niveau réseau, mais un bon paramétrage adaptatif des couches MAC et réseau peut mener à de meilleures performances en terme de débit et de consommation.

Le réseau que nous allons présenter dans ce qui suit se base sur une structure en grappe ou cluster.

La diffusion est un service fondamental dans les réseaux MANETs. La nature de la diffusion des transmissions sans fils, qui est que tous les voisins de l'émetteur reçoivent le paquet quand celui-ci émet un paquet, limite extrêmement l'extensibilité du réseau. Quand la taille du réseau augmente et le qu'il devient dense, une simple opération de diffusion peut déclencher d'importantes collisions et contentions de transmission pouvant amener à l'effondrement du réseau entier. Ceci est connu comme le problème de diffusion orageuse ou storm broadcast (Ni & al, 1999).

La structure en grappes est une implémentation simple du backbone qui contient seulement deux niveaux de structure hiérarchique. Le réseau est partitionné en groupes de grappes. Chaque grappe contient une tête de grappe (clusterhead) qui domine tous les autres nœuds. Deux têtes de grappe ne peuvent pas être voisines. Les passerelles ne sont pas des têtes de grappes et doivent avoir au minimum un voisin appartenant à une autre grappe

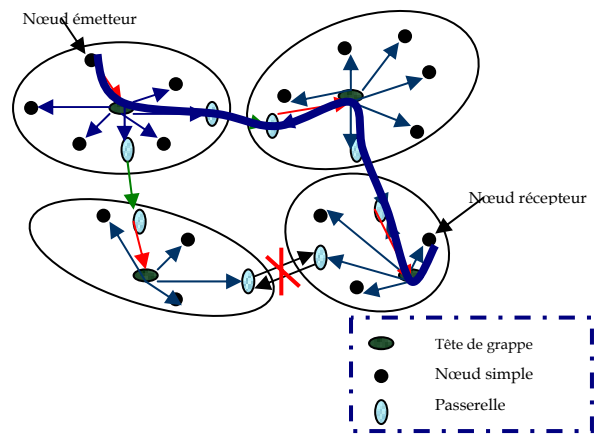

**Figure 2.** *Structure d'un MANET organisé en grappe*

L'acheminement de paquets dans le réseau se fait comme suit : le nœud émetteur, émet le paquet vers sa tête de grappe, qui se charge de le diffuser au niveau de la grappe. Au niveau des nœuds simples, s'ils ont reçus le paquet pour la première fois, ils vérifient si celui-ci leur est adressé, si c'est OUI ils exploitent le paquet, sinon il ne font rien

**4.2 Analyse des performances**

*4.2.1 Charge totale du réseau*

Dans cette section nous étudions le comportement du réseau du point de vue charge.

En d'autres termes en fonction de la charge du réseau, quelles sont les performances attendues ? Supposons qu'on n'a qu'un seul message dans le réseau, dans l'architecture que nous avons proposé, le message sera acheminer à travers les têtes de grappes et les passerelles jusqu'au destinataire. A ce niveau, nous n'allons pas trouver de problèmes qui soient propre à la couche MAC tel que les collisions. Mais si notre réseau contient une plus importante charge soit au niveau du nombre de nœuds, soit au niveau du nombre des messages circulant sur le réseau, là, plusieurs problèmes peuvent être se produire :

1- un grand nombre de collisions, entraînant de médiocres performances du réseau. Une solution à ce problème serait de se servir des mécanismes d'accès au média de la couche MAC IEEE 802.15.4, en définissant des niveaux de priorité aux messages et de véhiculer les messages prioritaires dans le segment CFP et les moins prioritaires sur le segment CAP. La tête de grappe serait le coordinateur du cluster (PAN).

2- Suite à un grand nombre de collisions ou de nœuds acheminant le paquet, le message peut arriver soit trop tard, soit ne pas arriver. Cela peut également arriver suite à un changement du cluster, dans lequel le message à déjà été diffusé, avant l'adhésion du nœud destinataire au nouveau cluster. Autrement dit lors de la diffusion du message le nœud destinataire était occupé à changer de cluster suite a un déplacement. Une solution serait de définir un temps de validité au niveau de chaque message, et ne plus l'acheminer à l'expiration de ce temps. Ou bien de définir une procédure d'adhésion à un cluster de telle sorte que si le clusterhead accepte un nouveau nœud dans son cluster, il vérifie s'il a diffusé un message juste avant d'accepter le nœud, si OUI et si il lui est adressé, il le lui transmet. Mais cette méthode oblige tous les clusterhead à garder tous les messages qu'ils diffusent pendant un laps de temps.

*4.2.2 Energie consommée*

Afin d'avoir une idée sur la marge de consommation en énergie des composants basé sur le IEEE 802.15.4, à titre d'exemple nous exposons au Tableau 1 les spécifications en consommation d'énergie du composant Chipcon CC2420, en fonction du mode de réception, de sommeil (IDLE), et d'arrêt.

| Power Down Mode (PD) | 20 | µA |
| Idle mode (IDLE) | 426 | µA |
| Current consumption, receive mode | 19.7 | mA |

**Tableau 1.** *Puissances consommées du 802.15.4 (Chipcon, 2004).*

Le ratio de puissance consommée est de 1 :46 en mode sommeil et 1 :985 en mode arrêt, par rapport à la consommation du circuit radio (émission/réception). Ceci montre l'intérêt d'affiner au mieux la partie active et inactive de la super trame.

La consommation en énergie est directement liée au trafic et à la charge dans le réseau, mais un bon routage peut économiser beaucoup d'énergie dans la mesure où il choisira le chemin qui impliquera le minimum de nœud dans l'acheminement du message tout en évitant qu'une grappe ne diffuse le même message plusieurs fois. Dans le réseau proposé plus haut et qui est basé essentiellement sur la diffusion le choix du chemin optimal est impossible. Il n'existe pas de table de routage. Dans ce cas précis, le seul remède est de définir une politique de diffusion intelligente et efficace.

Le nœud émetteur du message transmet le message à son clusterhead, puisque le clusterhead dispose d'une table contenant les identificateurs des nœuds de son cluster, il vérifie si le destinateur du message appartient a son cluster ou pas. Dans l'affirmative, il le lui transmet, sinon il le diffuse. Les passerelles se chargent de transmettre le message au cluster voisin, les passerelles réceptrices du message vérifient si elles ont déjà reçue le message ou pas. Si c'est oui elles ne font rien, sinon elle l'acheminent à leur clusterhead qui se charge de le diffuser après vérification du destinateur. Et ainsi de suite jusqu'à ce que le message soit arrivé à destination.

Dans ce type de routage, le problème est que malgré que le message soit délivré à son destinataire, les autres clusters continuent à le diffuser jusqu'à ce qu'il passe par tous les clusters, ou bien que sa date de validité soit expirée.

*Comment remédier à ce problème ?*

Nous n'avons pas de réponse à ce jour mais nous présentons les pistes de réflexion sur lesquelles nous travaillons au titre de conclusion sur cet article, et que nous résumons au paragraphe suivant.

## 5. Conclusion

Le routage est certainement le plus gros problème sur les MANETs, à traiter. Plusieurs équipes travaillent sur cet axe de recherche cet aspect. On pourra citer, entre autres, Interlab de l'université de Sherbrooke, le groupe IETF qui porte essentiellement sur le routage IP.

Un autre aspect de la consommation d'énergie est la consommation individuelle de chaque équipement. Nous avons présenté succinctement une analyse sur la consommation en énergie du IEEE 802.15.4, et les différents paramètres qui influent sur la consommation. A ce niveau on pourra adapter individuellement le comportement de ces paramètres de chaque cluster du réseau, en fonction du nombre de nœuds présent dans le cluster et du trafic (prioritaire CFP, non prioritaires CAP).

*5.1 Extensibilité du réseau*

Si le réseau défini plus haut devient très dense, un simple envoi de message peut amener au "crash" du réseau puisqu'un très grande diffusion peut amener à de forte collision et interférence entre différents équipements dans le réseau. Une des solutions serait de limiter le nombre de clusters et de faire en sorte que les zones de couverture des clusters ne se recouvrent pas. Ce qui veut dire que la portée d'émission de tous les nœuds soit juste suffisante pour couvrir les autres nœuds du cluster. Sans oublier que les passerelles des différents clusters doivent impérativement pouvoir communiquer entre elles. Chose qui est très difficile à mettre en œuvre dans une architecture mobile telle que les MANETs puisque les nœuds bougent et les clusterhead aussi. Sauf si l'on considère que les clusterhead sont des nœuds fixes, ce qui limiterait la mobilité du MANET, ce qui impliquera qu'il faudrait définir les clusterhead avant d'implémenter le réseau et l'extensibilité du réseau se verra réduite, mais dans cette architecture semi-fixe, il sera possible de définir des chemins de routage prédéterminés et assurer une délai maximal de transmission bout en bout pour des messages insérés dans la période sans contention PCF.

La norme IEEE 802.15.4 définit des portées de 10 et 100 m, l'utilisation d'un porté de 10 m pour les nœuds serait adéquate pour optimiser les interférences et les collisions, en plus la réduction de porté implique une économie de consommation d'énergie

*5.2 Débit*

En ce qui concerne le débit, la norme IEEE 802.15.4 définit un débit théorique de 115 Kbps, mais ce n'est que le débit théorique. Il est certain que la qualité des liaisons radio et la charge des protocoles : MAC, ainsi que les couches supérieures vont influer sur le débit, en plus si on commence à intégrer des fonctionnalité afin d'améliorer les performance du réseau ; ceci se traduirait directement par une baisse du débit effectif.

Par exemple le fait d'intégrer des échéances de validité sur les messages, ça va induire un plus grand nombre de paquets à transmettre et le débit effectif se verra diminuer.

La plupart du temps, les différentes décisions architecturales que nous serons amenés à prendre afin d'optimiser ces différents critères de configuration exposés ci-dessus, influeront d'une manière ou d'une autre sur le débit et la consommation énergétique. Notre travail actuel, consiste à trouver un modèle d'optimisation d'architecture en fonction de ces deux critères

## Références: